\documentclass[intlimits,twoside,a4paper]{article}
\usepackage[cp1251]{inputenc}

\usepackage{cmpj3}
\usepackage{bm}

\issue{2018}{21}{2}{23001}
\doinumber{10.5488/CMP.21.23001}

\title{Nonlinear dynamics of ferroelectrics with three-well local potential}
\author{R.~Yevych, M.~Medulych, Yu.~Vysochanskii}
\address{Institute for Solid State Physics and Chemistry, Uzhgorod National University,\\ 54 Voloshyn St., 88000 Uzhgorod, Ukraine}

\newcommand {\SPS}    {Sn$_2$P$_2$S$_6$ }

\date{Received March 22, 2018, in final form April 26, 2018}

\begin{document}

\maketitle

\begin{abstract}
For \SPS ferroelectrics, the appearance of spontaneous polarization is related to stereoactivity of tin cations and valence fluctuations of phosphorous cations. Here, the continuous phase transition and its behavior under pressure is determined by thee-well local potential and can be described in an anharmonic quantum oscillator model. For such a model, the spectrum of pseudospin fluctuations at different temperatures and pressures has been calculated and compared with the data of Raman spectroscopy. It was revealed that the ferroelectric lattice instability is related to several low energy optic modes.
\pacs 05.50.+q, 05.70.Fh, 63.70.+h, 77.84.-s
\end{abstract}

\section{Introduction}
For \SPS ferroelectrics, the second order phase transition at $T_0\approx 337$~K having a mixed displacive-order/disorder character takes place~\cite{vysoch2006}. The soft optic mode was observed by Raman scattering in ferroelectric phase, but temperature evolution of the soft mode spectrum is rather complicated --- several low energy optic branches linearly interact at heating, and the relaxational central peak grows in the vicinity of the phase transition temperature~\cite{bokot1997}. A complex transformation of \SPS lattice dynamics was also confirmed by inelastic neutron scattering~\cite{eijt1998} as well as by diffusive X-ray scattering~\cite{hlink2005}.

Calorimetric investigation~\cite{mori1988} reveals a high entropy of continuous ferroelectric transition in \SPS --- about 8.6~J$\cdot$K$^{-1}$mol$^{-1}$\, which is typical of an order/disorder transition. In addition, according to dielectric data for \SPS~\cite{grig1988}, the Curie-Weiss constant is approximately equal to $0.6\cdot10^5$~K which is typical of displacive transitions.

Crossover character (i.e., displacive-ordering) of the phase transition in \SPS was theoretically analyzed in a discrete model of a two-well (Ising-like) local potential~\cite{hlink1999}. It was found that the energy barrier between the wells and intercell interaction is equal to $U\approx4\cdot10^{-21}$~J. Closeness of these values well agrees with the mixed displacive-ordering character of the considered transition.

At further investigations using DFT techniques, it was found~\cite{rusch2007} that the local potential in \SPS ferroelectrics has a three-well shape. Peculiarities of the chemical bounding in this compound, which are related to  electron lone pair stereoactivity of Sn$^{2+}$ cations~\cite{rusch2007,glukh2012} and to valence fluctuations $\text{P}^{4+} + \text{P}^{4+}\rightarrow \text{P}^{3+} + \text{P}^{5+}$ inside (P$_2$S$_6$)$^{4-}$ anions, determine the three-well shape of the local potential for spontaneous polarization fluctuations~\cite{yevych20161}. Microscopic origin of ferroelectric lattice instability in \SPS can be described as the second-order Jahn-Teller (SOJT) effect~\cite{ber2013} and their thermodynamics can be considered within the framework of Blume-Emery-Griffiths model~\cite{blum1971_1,blum1971_2}. By taking into account the electronic correlations within the known Hubbard model, the presentation of Anderson's electron pairs flipping~\cite{andr1975_1,andr1975_2,andr1975_3} can be included into consideration of \SPS family ferroelectrics~\cite{yevych20161}. More generalized Hubbard-Holstein-like models treat electronic correlations together with their mixing with phonon excitations~\cite{fab1999}. In this approximation, having changed the local three-well potential by flattening the side wells, the calculated continuous phase transition temperature decreases and tricritical point (TCP) is reached~\cite{andr1975_1,andr1975_2,andr1975_3}. Below TCP temperature, the first order ferroelectric phase transition line further drops down to 0~K. In case of \SPS family ferroelectric crystals, such an evolution can be induced at substitution of tin for lead  or under hydrostatic compression~\cite{yevych20161}.

Electronic recharging and lattice instability can be presented as pseudospin fluctuations in anharmonic potential of a complicated shape. An anharmonic quantum oscillator (AQO) model was proposed earlier for a description of a temperature-pressure diagram of \SPS and of a temperature-composition diagram of (Pb$_y$Sn$_{1-y}$)$_2$P$_2$S$_6$ ferroelectric mixed crystals~\cite{yevych20161}.

The description of anharmonic dynamics for ferroelectric lattices is a very complicated task. Early anharmonic splitting of the soft optic mode was theoretically predicted~\cite{stas1997} and experimentally observed~\cite{grig1984} for SbSJ ferroelectrics with a two-well local potential. In this paper we use the AQO model having a three-well local potential for the purpose of analysing the \SPS ferroelectric dynamics at variation of temperature and pressure.

\section{Short description of a model}
In the present work we used the AQO model for studying the dynamic and dielectric properties of \SPS ferroelectric crystal. Such a model was successfully used earlier for investigations of structural phase transitions of  model and real systems~\cite{gill1974,flock1989,stas1997,bakk1993}. In the AQO model, the real crystal lattice is represented as a system of one-dimensional interacting quantum anharmonic oscillators. In this approach, the model Hamiltonian is as follows:
\begin{equation}\label{eq1}
    H=\sum_i\left[T(p_i)+V(x_i)\right]-\sum_{ij}J_{ij}x_ix_j\,,
\end{equation}
where  $T(p_i)$ and $V(x_i)$ are kinetic and local potential energies, respectively, as functions of the momentum $p_i$ and position $x_i$ operators of the $i$-th oscillator; $J_{ij}$ are the coupling constants between $i$-th and $j$-th oscillators. In order to simplify the equation~(\ref{eq1}), a mean-field approach was used. It is assumed that each oscillator is influenced by a self-consistent linear symmetry-breaking field which depends on the average displacement of a system of  oscillators. Thus, the last term in Hamiltonian~(\ref{eq1}) which describes an interaction can be replaced by
\begin{equation}\label{eq2}
    \sum_{ij}J_{ij}x_ix_j\rightarrow \sum_{i}Jx_i\langle x\rangle,
\end{equation}
\noindent where $J$ is the coupling constant between the field and oscillators, $\langle x\rangle$ is the average displacement of oscillator. This suggestion makes it possible to represent a model Hamiltonian as a sum of one-particle non-interacting Hamiltonians:
\begin{align}\label{eq3}
    H&=\sum_iH_{i}^{\textrm{eff}},\nonumber\\
    H_{i}^{\textrm{eff}}&=T(p_i)+V(x_i)-Jx_i\langle x\rangle.
\end{align}
Having self-consistently solved the Schr\"{o}dinger equation with an effective Hamiltonian (\ref{eq3}), it is easy to obtain a full set of eigen-energies $\{E_n\}$ and wave functions $\{\Psi_n(x)\}$. The average displacement $\langle x\rangle$ is determined at each temperature $T$ by:
\begin{eqnarray}\label{eq4}
    \langle x\rangle=\sum_n P_n\langle x_n\rangle, \qquad
    P_n=\frac{\re^{-E_n/kT}}{\sum_n\re^{-E_n/kT}}\,,\qquad \langle x_n\rangle=\frac{\int \Psi_n^*x\Psi_n\rd x}{\int \Psi_n^*\Psi_n\rd x}\,,
\end{eqnarray}
where $P_n$ is a probability to find the oscillator at the $n$-th level with the energy $E_n$, $\langle x_n\rangle$ is the average displacement for the $n$-th level, $k$ is the Boltzmann constant. On numerical calculations, there was used a matrix representation form for position and momentum operators \cite{korch}. Eigenfunctions of anharmonic oscillator have been found as a series of eigenfunctions of a harmonic oscillator. The lowest 200 energy levels were accounted for in the present work to obtain a satisfying accuracy on calculations.

To study the influence of an external field $E$ on the system under consideration, the term proportional to $Ex_i$ should be included in the equation~(\ref{eq3}). To investigate the phonon-like excitations within the framework of AQO model, the phonon spectral function was calculated, given by~\cite{oshiba}
\begin{equation}\label{eq5}
\rho_{\textrm{ph}}(\omega)=\sum_{ij}A_{ij}\delta(\omega+E_i-E_j).
\end{equation}
The spectral weight $A_{ij}$ is given by
\begin{equation}\label{eq6}
A_{ij}=\frac{\re^{-E_i/kT}-\re^{-E_j/kT}}{\sum_i\re^{-E_i/kT}}\left|\int\Psi_i^*x\Psi_j\rd x\right|^2. \nonumber
\end{equation}
To simulate a broadening, each spectral line has been replaced by Lorentzian profiles with the same half-width of 3~cm$^{-1}$ and intensities proportional to their spectral weight.

Since an order parameter is proportional to the average displacement $\langle x\rangle$, one can investigate a different type of a phase diagram by changing the temperature, a coupling constant or the shape of the local potential. These simulations may correspond to the applied pressure or chemical substitution, for example, in real systems.

\section{Discussion of results}
\begin{figure}[!b]
\centering
   \includegraphics[width=6cm]{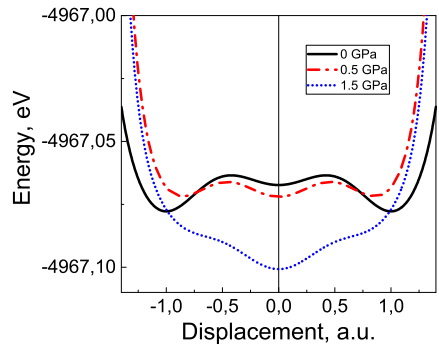}%
  \caption{(Colour online) The shape of the local potential for \SPS crystal at different pressures~\cite{rusch2016}.}\label{fig1}
\end{figure}

\begin{figure}[!t]
\centering
   \includegraphics[width=5.5cm]{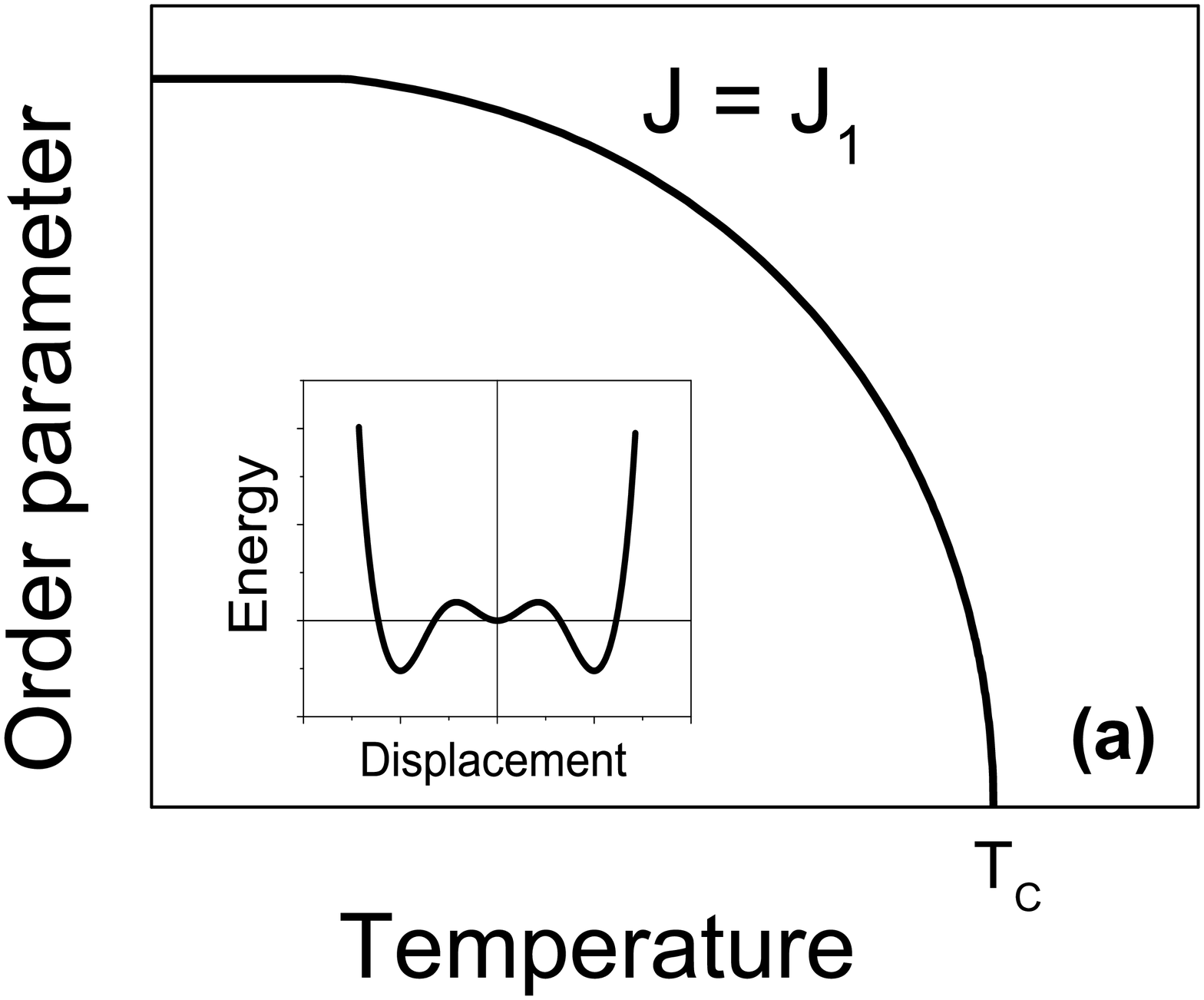} \qquad
   \includegraphics[width=5.5cm]{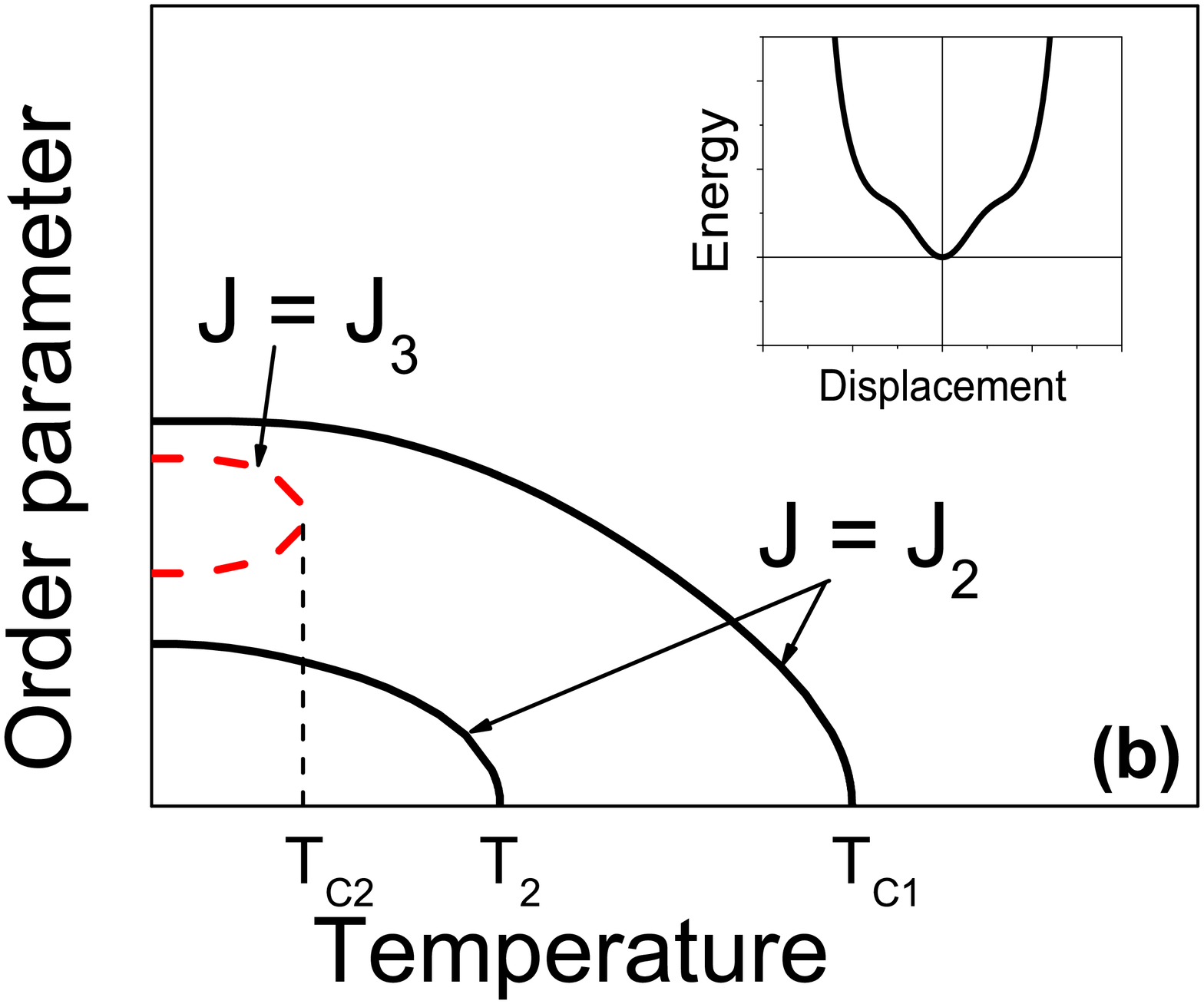}
  \caption{(Colour online) Typical temperature dependencies of the order parameter at a different shape of three-well local potential (on the inset): (a) --- continuous transition at $T=T_{\text C}$ and $J_1=9$~J/m$^2$; (b) --- continuous transition at $J_2=11$~J/m$^2$ and $T=T_{\textrm C1}$ (metastable solution exists at $T\leqslant T_2$), discontinuous transition at $J_3=9$~J/m$^2$ and $T=T_{\textrm C2}$ (metastable and stable solutions coexist in the same temperature range).}\label{fig2}
\end{figure}

\begin{figure}[!t]
\centering
   \includegraphics[width=8cm]{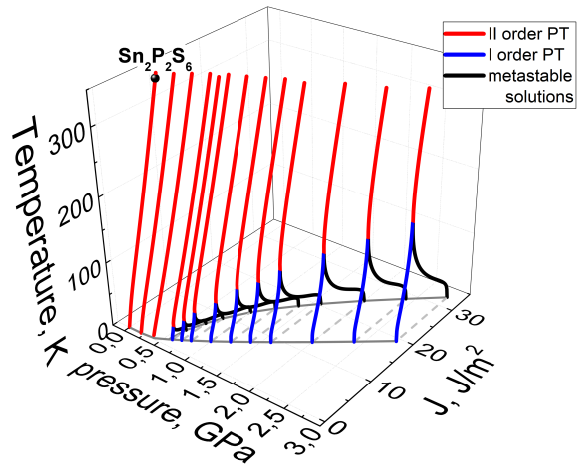}%
  \caption{(Colour online) The temperature-pressure-coupling constant phase diagram of the systems having a local three-well potential~\cite{yevych20161}. The position of \SPS crystal on the diagram is marked with a black circle.}\label{fig3}
\end{figure}

In the present work we use a three-well local potential for order parameter fluctuations of \SPS crystal~\cite{rusch2007}. The evolution of this potential with pressure is presented in figure~\ref{fig1}.  As can be seen in figure~\ref{fig1},  transformation of the shape of the potential under compression has a interesting feature: at a pressure equal to the characteristic one $p_{\textrm c}\approx0.5$~GPa, all three wells are of the same depth (at $p<p_{\textrm c}$, the side-wells are deeper than the central one, and at  $p>p_{\textrm c}$, the central well becomes the deepest). Typical temperature dependencies of the order parameter depending on the shape of the potential and coupling constant $J$ are shown in figure~\ref{fig2}. It should be noted that the coupling constant has mainly an effect on temperature of the phase transition. However, the shape of the potential determines a possibility of realizing a certain type of transition in the system. Based on this potential, the AQO model describes the temperature-pressure-coupling constant phase diagram for \SPS-like crystals as shown in figure~\ref{fig3}~\cite{yevych20161}. Here, the characteristic pressure $p_{\textrm c}$ is a lower bound of a region on the pressure scale where the first-order phase transitions and metastable solutions (with higher energy of a ground state) can be realized. The case of \SPS crystal corresponds to the value of the coupling constant about 8.82~J/m$^2$ on the represented diagram.

The fluctuations of spontaneous polarization are determined by three values of pseudospin  (``$-1$'', ``0'', and ``$+1$''), and the thermodynamics of such a system is described by BEG model~\cite{blum1971_1,blum1971_2}. The pseudospin is composed of ion and electron components which are examined in Hubbard-Holstein models~\cite{fab1999}. Corresponding contributions to spontaneous polarization relate to normal coordinates of polar phonons with $B_u$ symmetry and to a recharging within SnPS$_3$ structural groups~\cite{yevych20161}. The ion component of spontaneous polarization mainly relates to low-frequency phonon modes --- the translational vibrations of a crystal lattice with frequencies below 150~cm$^{-1}$. Being determined by charge disproportion, the electron recharging can be generally associated with internal vibrations of (P$_2$S$_6$)$^{4-}$ anion structural groups, that is, can be realized through the changes of P-P and P-S bond length, that belong to 375--600~cm$^{-1}$ frequency range~\cite{vysoch2006}. It can be supposed that variation of the energy levels of a quantum anharmonic oscillator qualitatively reflects the temperature evolution of nonlinear dynamics of a ferroelectric system having a local many-well potential.
\begin{figure}[!t]
\centering
   \includegraphics[width=5cm]{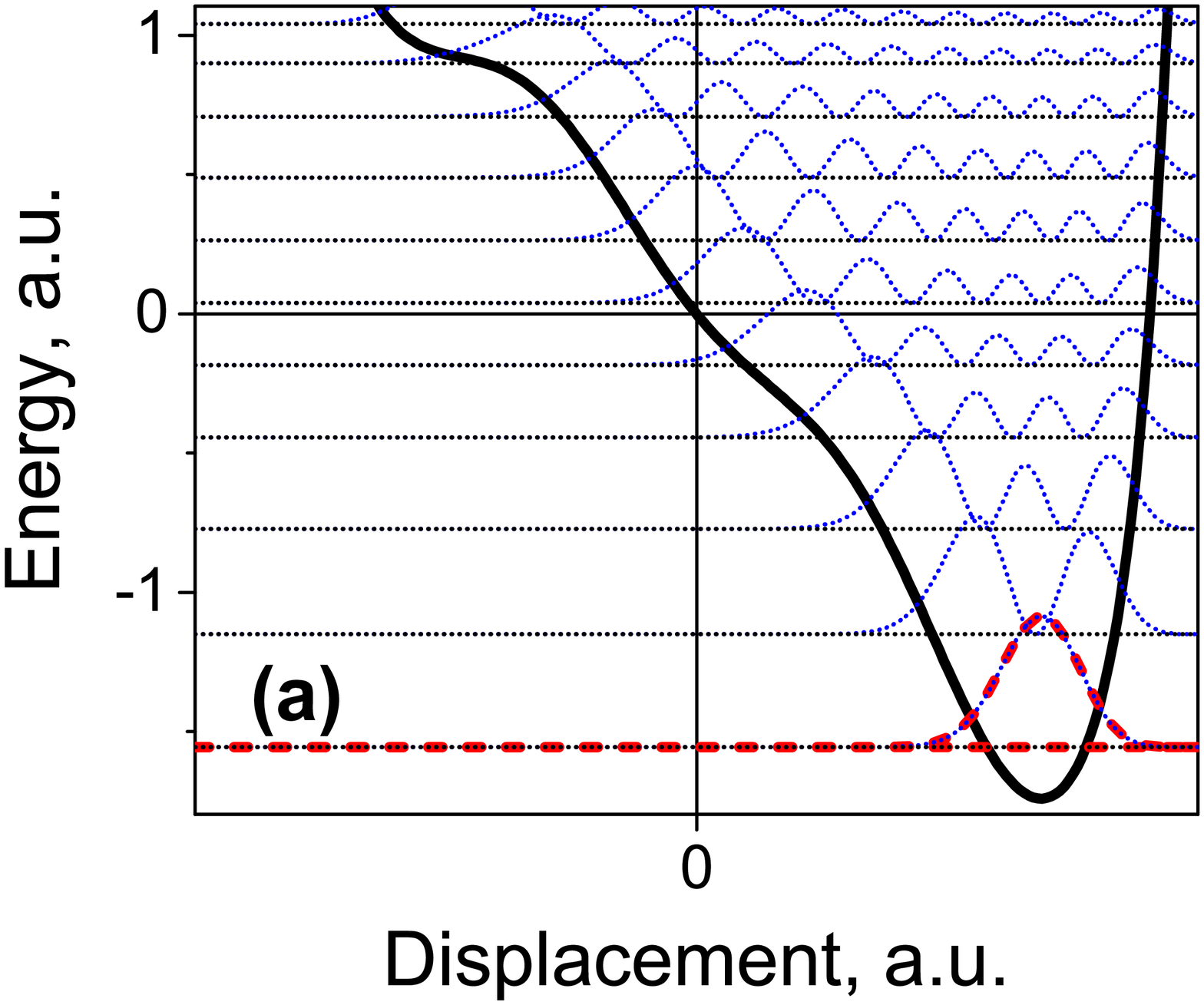}%
   \includegraphics[width=5cm]{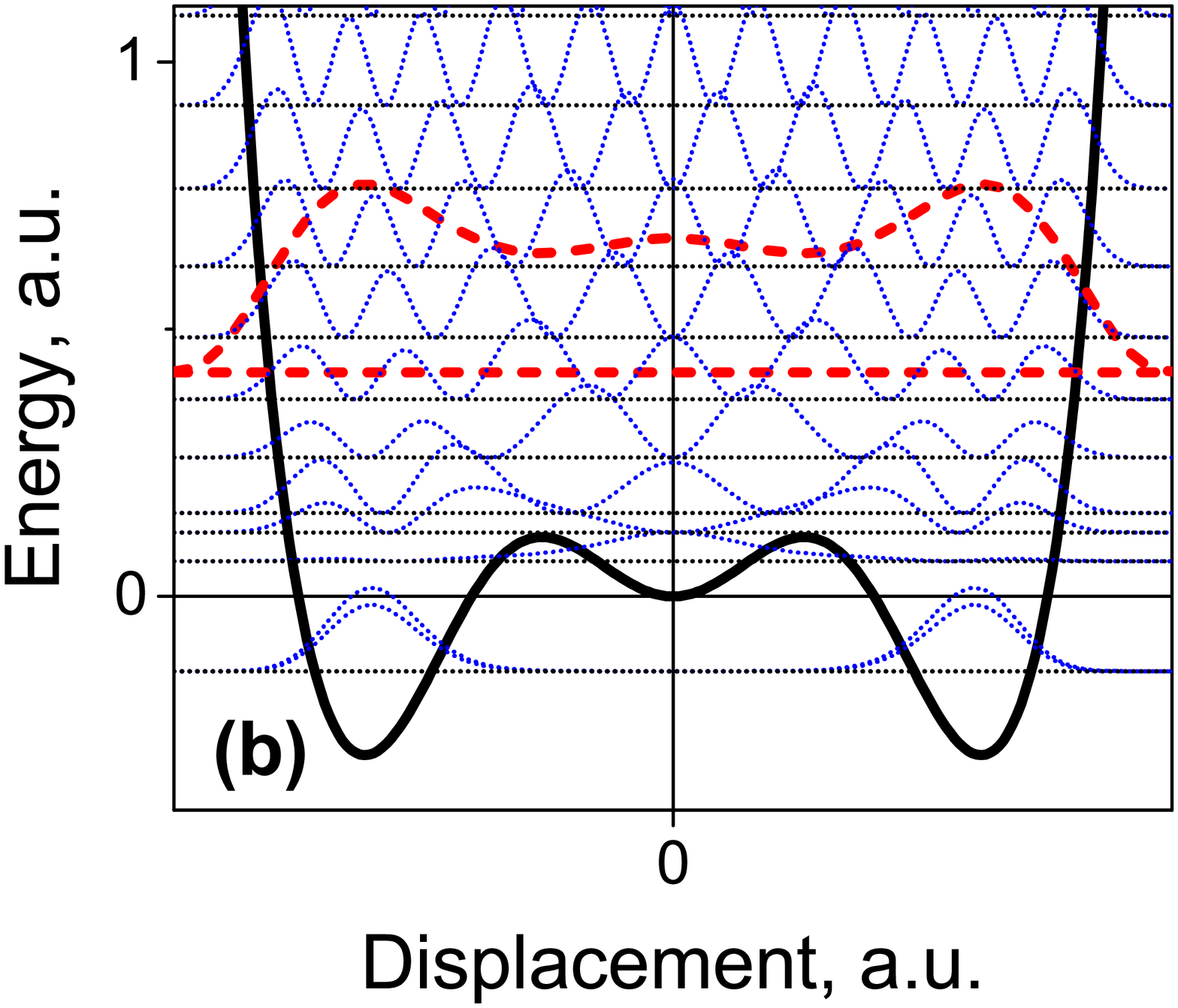}%
   \includegraphics[width=5cm]{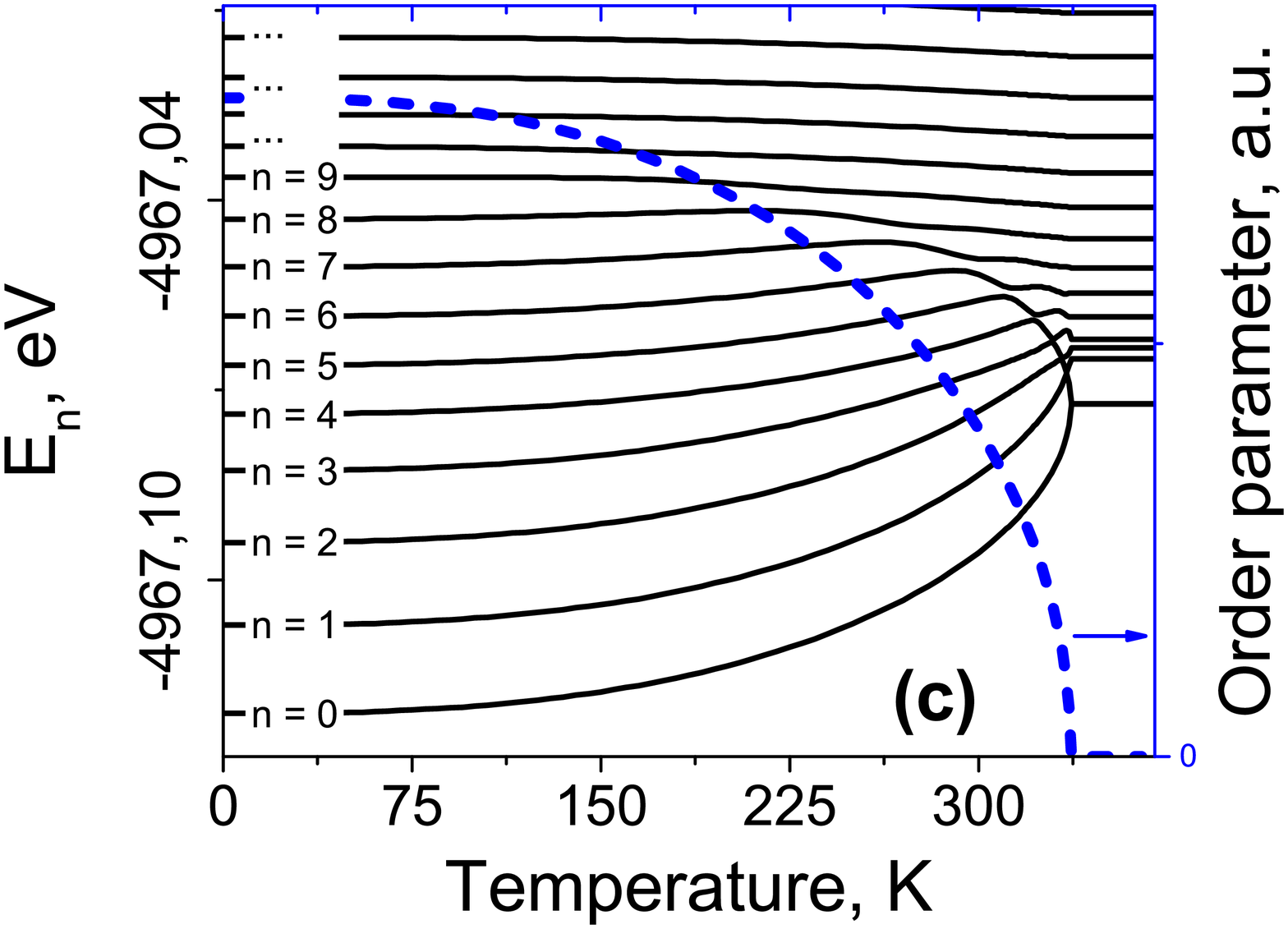}
  \caption{(Colour online) Local three-well potential at $T=0$~K (a) and $T=350$~K (b) (thick black line) and their energy levels with corresponding squared wave functions (thin black and blue lines). According to probabilities by equation~(\ref{eq4}) at a given temperature, the average energy levels and wave functions were calculated (dashed red lines). (c) --- temperature dependence of a few low-energy levels (black solid lines) and order parameter (blue dashed line).}\label{fig4}
\end{figure}
Being calculated within the mean-field approach, the temperature evolution of the energy spectrum (see figure~\ref{fig4}) correlates with the temperature dependence of spontaneous polarization which disappears at heating and is accompanied by a transition from ferroelectric phase to the paraelectric one. Theoretical pressure-temperature-coupling constant phase diagram (see figure~\ref{fig3}) shows that the temperature of phase transition drops down to 0~K at a pressure about 2.2~GPa for \SPS crystal. Moreover, tricritical point is indicated near the temperature of 250~K and within 0.4--0.6~GPa pressure range. Using AQO model it is possible to analyze the temperature transformation of phonon spectra associated with spontaneous polarization for \SPS crystal.

\begin{figure}[!t]
\centering
   \includegraphics[width=6cm]{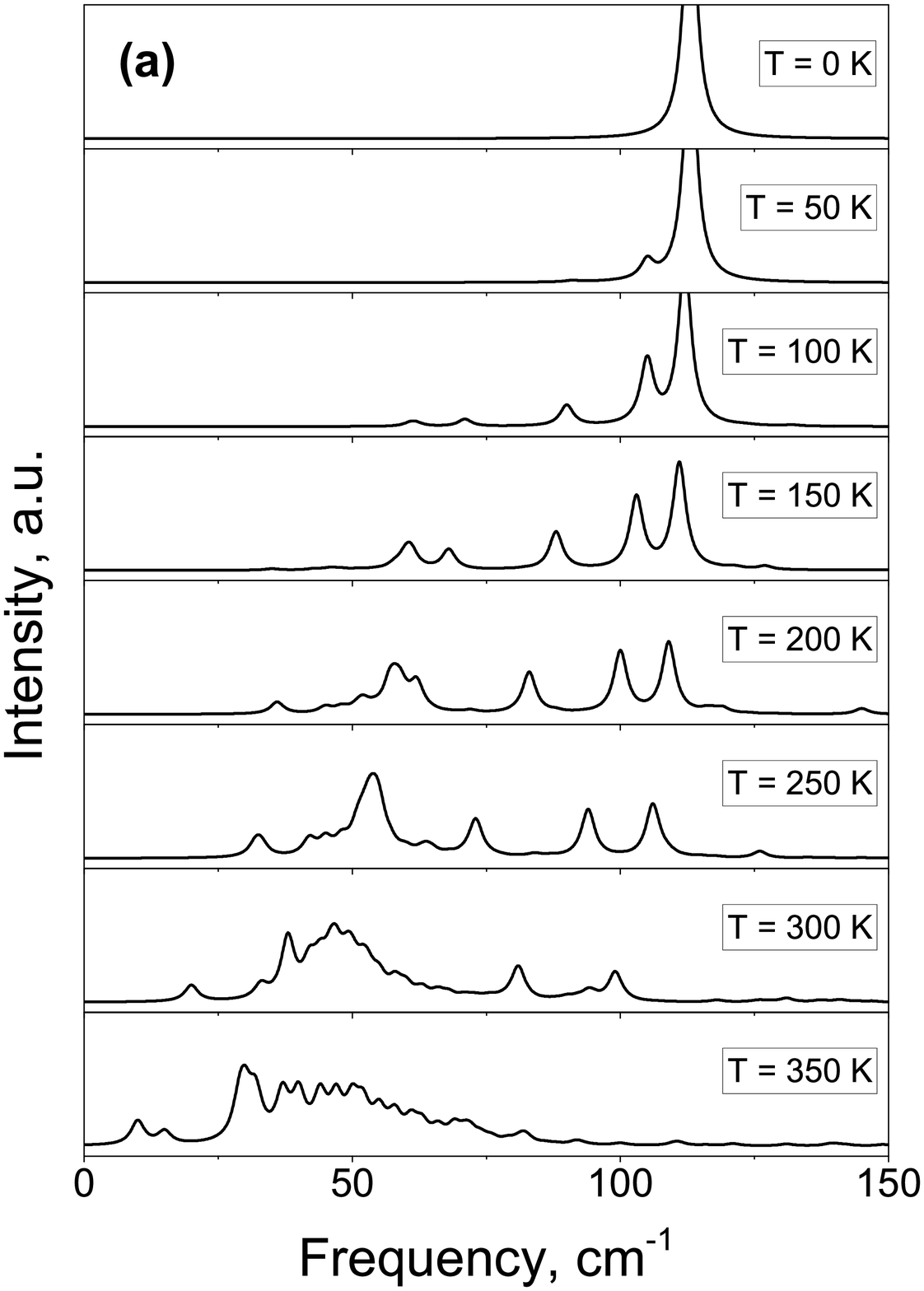} \qquad
   \includegraphics[width=6cm]{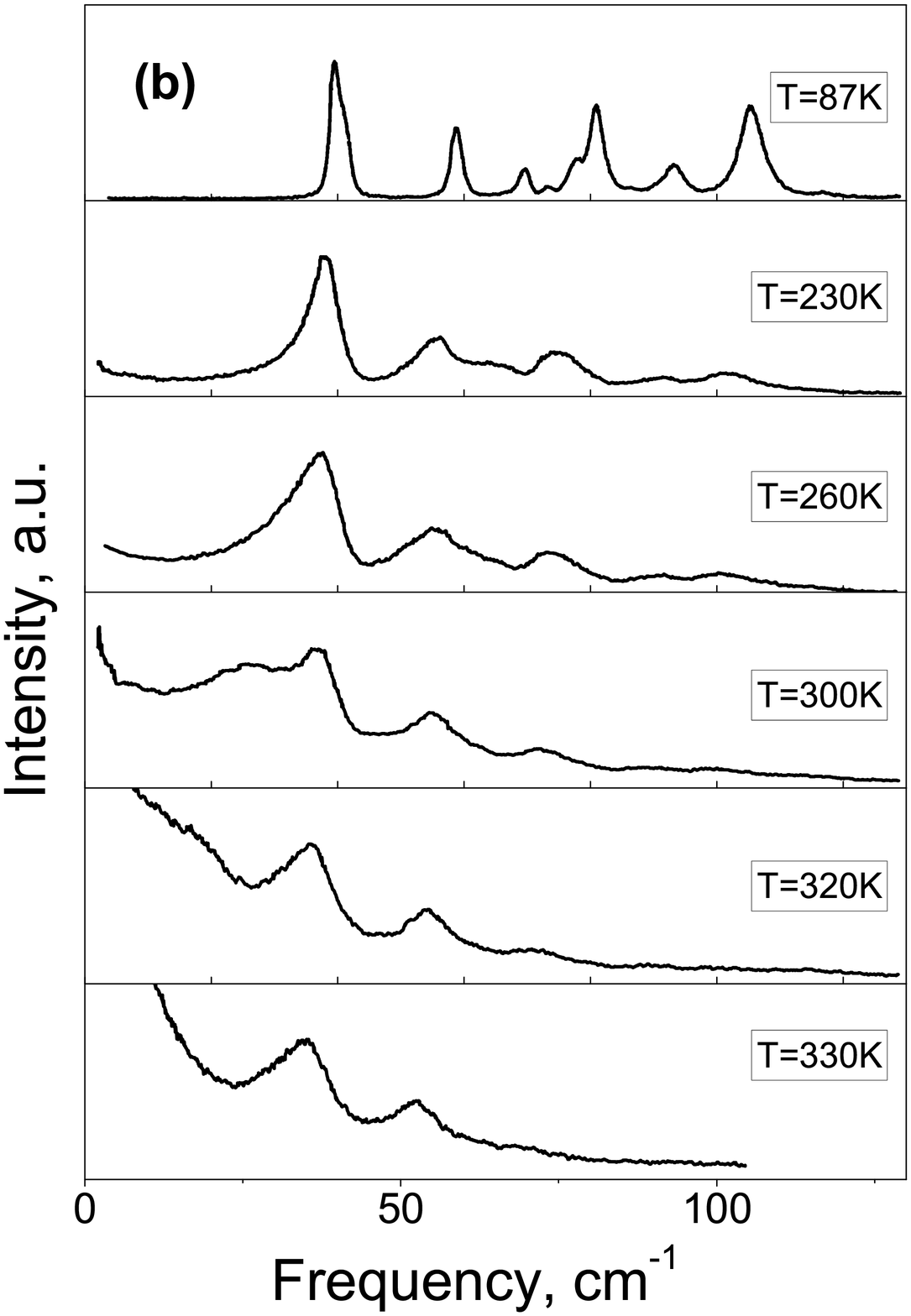}
  \caption{The calculated spectra for a three-well potential (a) and the observed Raman spectra (b) of \SPS at different temperatures for a normal pressure.}\label{fig5}
\end{figure}

The temperature transformation of the spectrum corresponding to transitions between energy levels of quantum oscillator having a three-well potential at an ambient pressure is shown in figure~\ref{fig5}~(a). At low temperatures, the transitions are excited between the levels with the energy difference about 120~cm$^{-1}$. At a temperature increase, the probability of an occupancy at higher energy non-equidistant levels grows which becomes evident due to the appearance of spectral lines of a lower energy. Approaching the phase transition point ($T_0=337$~K), the spectral weight passes to a group of overlapping lines in the frequency range of 25--80~cm$^{-1}$. A separate line becomes apparent about 12~cm$^{-1}$ which can be related to the experimentally observed soft optic mode. There are low frequency (from 10 to 30~cm$^{-1}$) spectral lines in the paraelectric phase that do not show a significant behavior while temperature increases.

\begin{figure}[!t]
\centering
   \includegraphics[width=6.5cm]{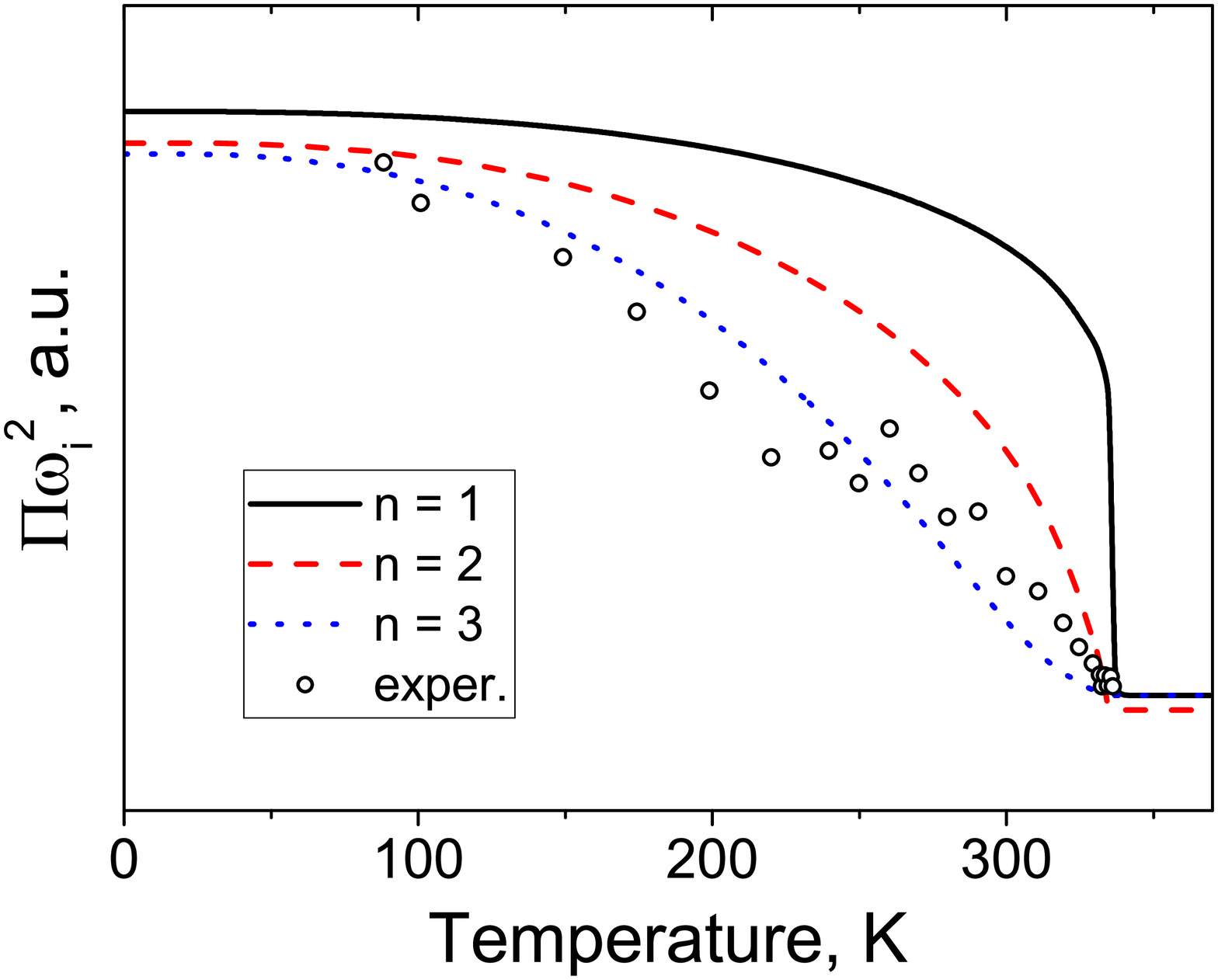}
  \caption{(Colour online) The calculated temperature dependencies of the product of squared frequencies for $n$ lower modes (lines). Experimental data from work~\cite{bokot1997} are shown by open circles.}\label{fig5c}
\end{figure}

The calculated temperature evolution of the energy spectrum for a system of anharmonic quantum oscillators generally agrees with the tendency of temperature changes in experimentally observed Raman spectra in \SPS crystal [see figure~\ref{fig5}~(b)]. At heating in a ferroelectric phase, low energy spectral lines slightly decrease its frequency and significantly change the width and the shape asymmetry. The wide spectral line which is identified as a soft optic mode~\cite{vysoch2006,bokot1997} appears below 40~cm$^{-1}$. It spreads with the temperature increase and its maximum shifts to lower frequencies. In addition, there appears a central peak of quasielastic light scattering which can be related to the relaxation dynamics of pseudospins between central and side wells of the local potential. According to the experimental Raman spectra temperature evolution \cite{bokot1997} in the ferroelectric phase of \SPS crystal, the product of squared frequencies for several lowest energy optic modes is the most satisfactory only regarding the expected linear temperature dependence at heating  to the second order phase transition temperature. Such a manifestation of the lattice instability related to the structural transition agrees with the temperature dependence of the energy distance between the levels calculated in the AQO model (see figure~\ref{fig5c}).

\begin{figure}[!b]
\vspace{-3mm}
\centering
   \includegraphics[width=6cm]{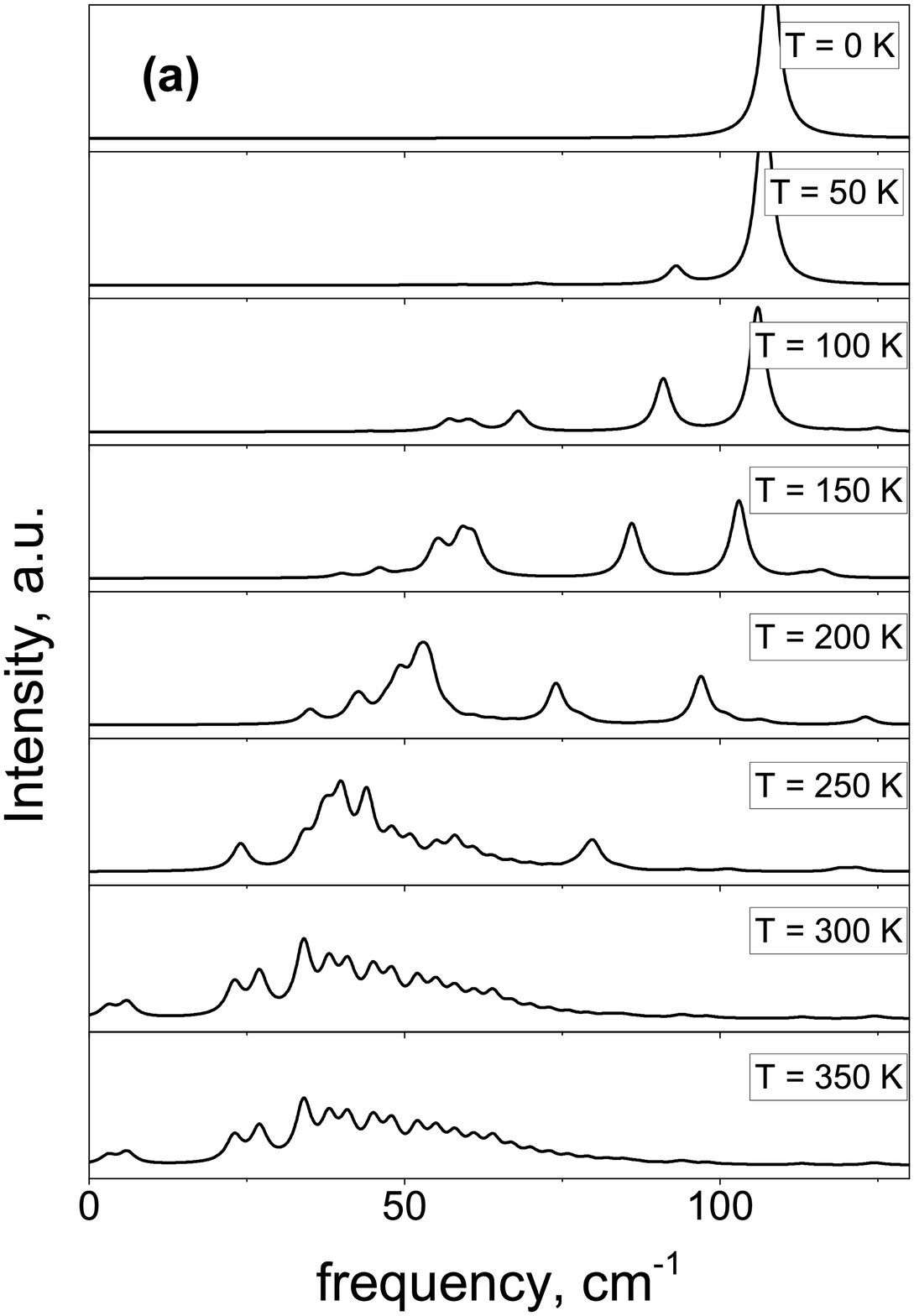} \qquad
   \includegraphics[width=6cm]{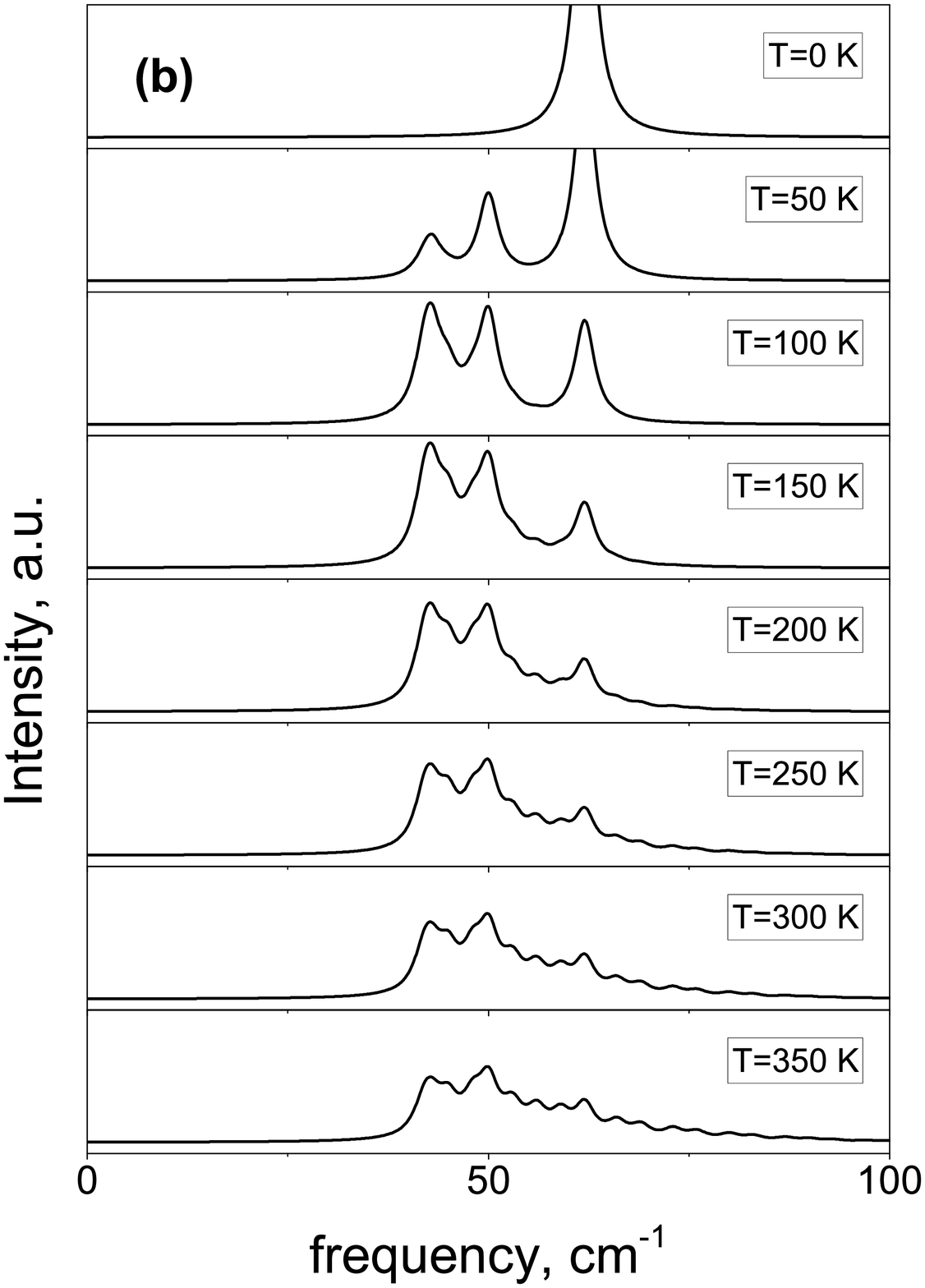}
  \caption{The calculated spectra for a three-well potential at different temperatures for pressures: (a)~---~0.4~GPa, (b) --- 2.2~GPa.}\label{fig6}
\end{figure}

\begin{figure}[!t]
\vspace{-6mm}
\centering
   \includegraphics[width=6cm]{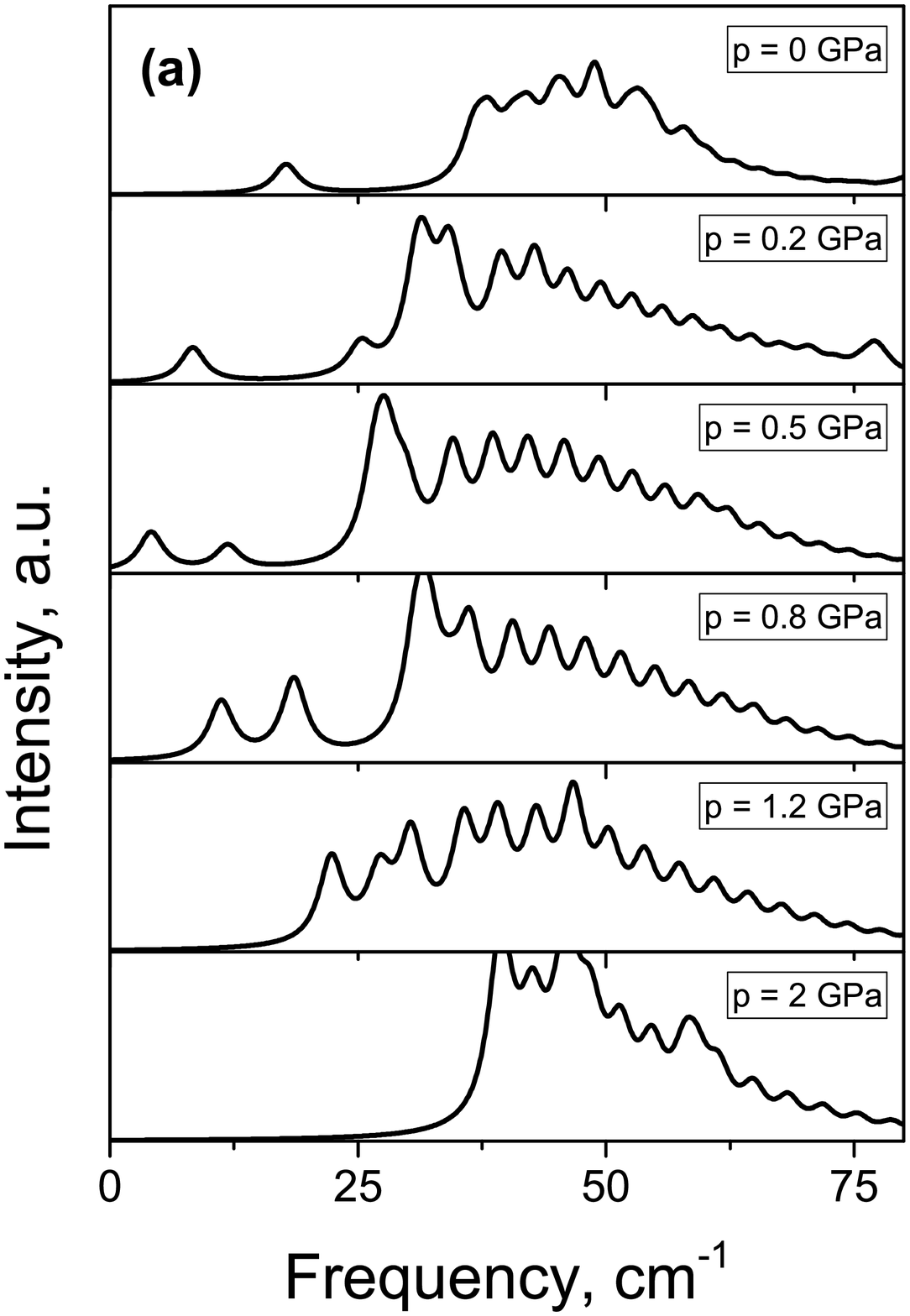} \qquad
   \includegraphics[width=6cm]{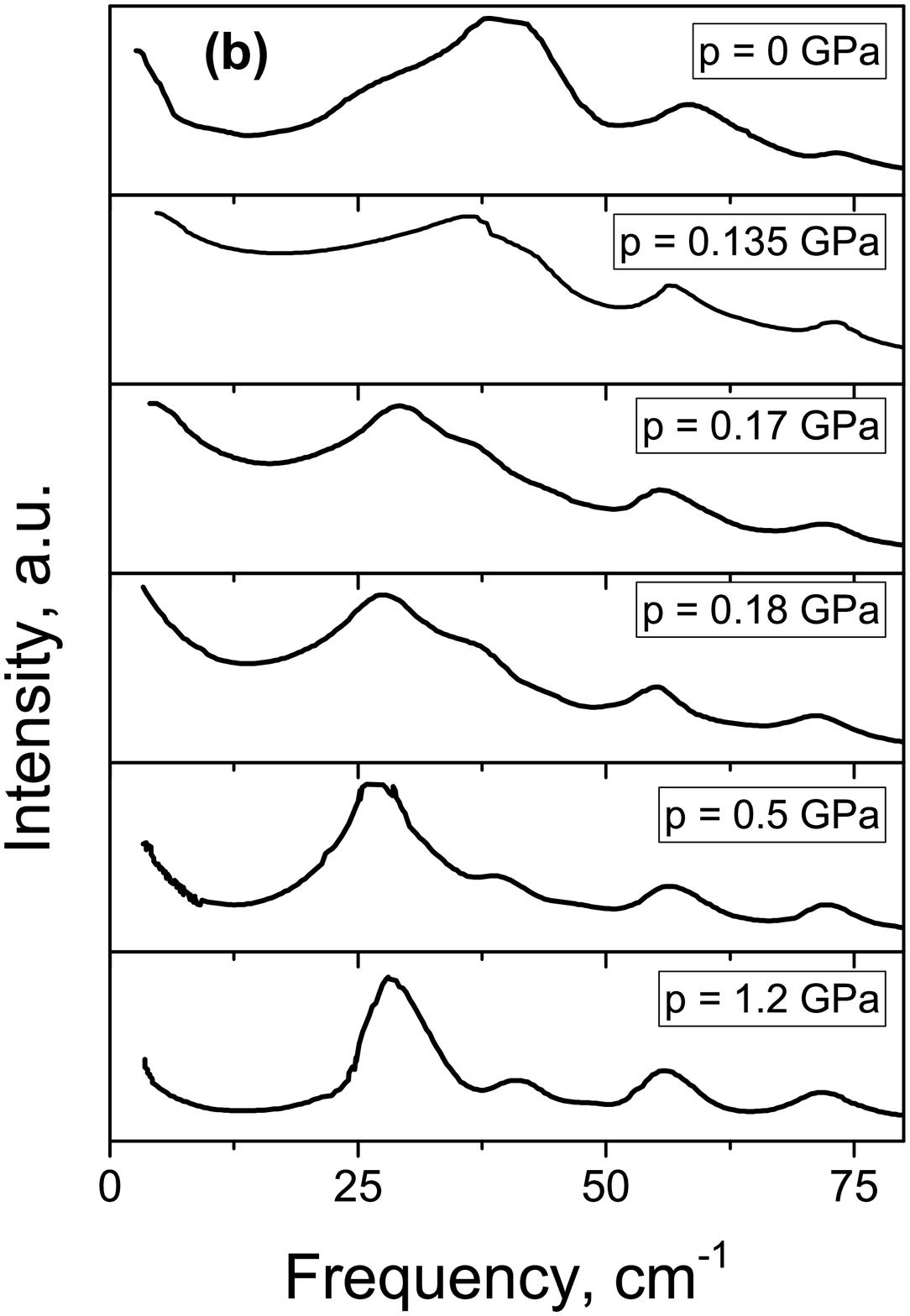}
  \caption{The calculated spectra for a three-well potential (a) and the observed Raman spectra~\cite{vysoch2006} (b) of \SPS at different pressures for room temperature.}\label{fig7}
\end{figure}

It is interesting to analyse the influence of pressure on the calculated temperature transformation of the energy spectrum of a system of AQOs. At a tricritical pressure, a broad peak caused by the merging of a group of spectral lines appears below 50~cm$^{-1}$ at a lower temperature --- about 250~K (see figure~\ref{fig6}). However, the lowest energy band is found at a frequency 25~cm$^{-1}$ here. In the paraelectric phase at $T=300$~K, the lowest frequency line is located near 5~cm$^{-1}$ which indicates a rapid temperature evolution of the spectrum of pseudospin excitations at heating in the vicinity of tricritical point. Under the pressure of 1~GPa and at temperature of 200~K in the paraelectric phase, the lowest band in the spectrum lies at 20~cm$^{-1}$. By applying a pressure of 1.5~GPa, the paraelectric phase becomes stable at cooling to low temperatures (see figure~\ref{fig3}). In this case, a group of lines below 50~cm$^{-1}$ manifests itself already at $T=50$~K. At a further temperature increase, there is almost a continuous spectrum in the range from 25 to 75~cm$^{-1}$. While the pressure grows up to 2.2~GPa, the calculated spectrum contains a more contrast set of lines, which is shifted to higher frequencies (see figure~\ref{fig6}) in accordance with the experimental data regarding the pressure influence on the Raman spectra in a paraelectric phase of \SPS~\cite{vysoch2006}. Besides, the pressure transformation of experimentally observed Raman spectra at room temperature qualitatively agreers with the calculated one (see figure~\ref{fig7}). Approaching the critical point from both low or high pressures, the low energy spectral lines obviously decrease their frequencies.

\section{Conclusions}
For \SPS proper uniaxial ferroelectrics with a local three-well potential for spontaneous polarization fluctuations, an anharmonic quantum oscillator model has been developed and used for a description of a lattice instability related to a ferroelectric phase transition. The calculated temperature dependence of pseudospin fluctuations agrees with the observed changes of Raman spectra at heating in a ferroelectric phase. The hydrostatic pressure influence on the local potential shape is clearly observed by the variation of the spectra of pseudospin fluctuations. The lattice instability related to a continuous phase transition is manifested as a linear temperature dependence for a product of squared frequencies of several low energy optical modes. A complicated ``soft mode'' spectrum is a characteristic peculiarity of a mixed displacive-order/disorder transition with a rearrangement of crystal chemical bounding by Sn$^{2+}$ cation stereoactivity and by the valence fluctuations $\text{P}^{4+} + \text{P}^{4+}\rightarrow \text{P}^{3+} + \text{P}^{5+}$ for phosphorous cations.

\section*{Acknowledgements}
Authors acknowledge helpful discussions with Prof. Ihor V. Stasyuk.

\ukrainianpart

\title{Нелінійна динаміка сегнетоелектриків із триямним потенціалом}
\author{Р.~Євич, М.~Медулич, Ю.~Височанський}
\address{
Науково-дослідний інститут фізики і хімії твердого тіла, Ужгородський національний університет,\\ вул. Волошина, 54, 88000 Ужгород, Україна
}

\makeukrtitle

\begin{abstract}
\tolerance=3000%
Для сегнетоелектрика \SPS поява спонтанної поляризації пов'язана зі стереоактивністю катіонів олова та валентними флуктуаціями катіонів фосфору. Неперервний фазовий перехід та його поведінка з тиском визначається локальним триямним потенціалом і описується в моделі ангармонічного квантового осцилятора. Для такої моделі розраховані спектри флуктуацій псевдоспіна при різних температурах і тисках, які порівнюються з даними раманівської спектроскопії. Показано, що нестійкість сегнетоелектричної ґратки пов'язана з декількома низькоенергетичними оптичними модами.
\keywords фазові діаграми, сегнетоелектрик, ґраткові моделі в статистичній фізиці

\end{abstract}

\end{document}